\batchmode
\documentstyle[12pt]{article}
\newcommand{\nc}{\newcommand}
\nc{\renc}{\renewcommand}

\nc{\half}{{\textstyle{1\over2}}}
\nc{\etal}{\mbox{\it et al. }}
\nc{\ie}{{\it i.e.}}
\nc{\eg}{{\it e.g.}}

\renc{\thefootnote}{\arabic{footnote}}
\nc{\capt}[1]{{\bf Figure.} {\small\sl #1}}

\nc{\eqs}[2]{\mbox{Eqs.~(\ref{#1},\,\ref{#2})}}
\nc{\eq}[1]{\mbox{Eq.~(\ref{#1})}}

\nc{\figs}[2]{\mbox{Figs.~(\ref{#1},\,\ref{#2})}}
\nc{\fig}[1]{\mbox{Fig~.(\ref{#1})}}

\nc{\tag}[1]{\label{#1} \marginpar{{\footnotesize #1}}}
\nc{\mtag}[1]{\label{#1} \mbox{\marginpar{{\footnotesize #1}}}}
\renc{\baselinestretch}{1.5}
\newlength{\overeqskip}
\newlength{\undereqskip}
\nc{\be}[1]{\begin{equation} \mbox{$\label{#1}$}}
\nc{\bea}[1]{\begin{eqnarray} \mbox{$\label{#1}$}}
\nc{\Section}[2]{\section{#2}\label{#1}}
\nc{\Bibitem}[1]{\bibitem{#1}}
\nc{\Label}[1]{\label{#1}}

\nc{\eea}{\vspace{\undereqskip}\end{eqnarray}}
\nc{\ee}{\vspace{\undereqskip}\end{equation}}
\nc{\bdm}{\begin{displaymath}}
\nc{\edm}{\end{displaymath}}
\nc{\dpsty}{\displaystyle}
\nc{\bc}{\begin{center}}
\nc{\ec}{\end{center}}
\nc{\ba}{\begin{array}}
\nc{\ea}{\end{array}}
\nc{\bab}{\begin{abstract}}
\nc{\eab}{\end{abstract}}
\nc{\btab}{\begin{tabular}}
\nc{\etab}{\end{tabular}}
\nc{\bit}{\begin{itemize}}
\nc{\eit}{\end{itemize}}
\nc{\ben}{\begin{enumerate}}
\nc{\een}{\end{enumerate}}
\nc{\bfig}{\begin{figure}}
\nc{\efig}{\end{figure}}
\nc{\arreq}{&\!=\!&}
\nc{\arrmi}{&\!-\!&}
\nc{\arrpl}{&\!+\!&}
\nc{\arrap}{&\!\!\!\approx\!\!\!&}
\nc{\non}{\nonumber\\*}
\nc{\align}{\!\!\!\!\!\!\!\!&&}

\def\lsim{\; \raise0.3ex\hbox{$<$\kern-0.75em
      \raise-1.1ex\hbox{$\sim$}}\; }
\def\gsim{\; \raise0.3ex\hbox{$>$\kern-0.75em
      \raise-1.1ex\hbox{$\sim$}}\; }
\nc{\DOT}{\hspace{-0.08in}{\bf .}\hspace{0.1in}}
\nc{\Laada}{\hbox {$\sqcap$ \kern -1em $\sqcup$}}
\nc\loota{{\scriptstyle\sqcap\kern-0.55em\hbox{$\scriptstyle\sqcup$}}}
\nc\Loota{{\sqcap\kern-0.65em\hbox{$\sqcup$}}}
\nc\laada{\Loota}
\nc{\qed}{\hskip 3em \hbox{\BOX} \vskip 2ex}

\nc{\real}{{\rm I \! R}}
\nc{\Z}{{\sf Z \!\!\! Z}}
\nc{\complex}{{\rm C\!\!\! {\sf I}\,\,}}
\def\bigid{\leavevmode\hbox{\small1\kern-3.8pt\normalsize1}}
\def\id{\leavevmode\hbox{\small1\kern-3.3pt\normalsize1}}
\nc{\slask}{\!\!\!/}
\nc{\bis}{{\prime\prime}}
\nc{\pa}{\partial}
\nc{\na}{\nabla}
\nc{\ra}{\rangle}
\nc{\la}{\langle}
\nc{\goto}{\rightarrow}
\nc{\swap}{\leftrightarrow}

\nc{\EE}[1]{ \mbox{$\cdot10^{#1}$} }
\nc{\abs}[1]{\left|#1\right|}
\nc{\at}[2]{\left.#1\right|_{#2}}
\nc{\norm}[1]{\|#1\|}
\nc{\abscut}[2]{\Abs{#1}_{\scriptscriptstyle#2}}
\nc{\vek}[1]{{\rm\bf #1}}
\nc{\integral}[2]{\int\limits_{#1}^{#2}}
\nc{\inv}[1]{\frac{1}{#1}}
\nc{\dd}[2]{{{\partial #1}\over{\partial #2}}}
\nc{\ddd}[2]{{{{\partial}^2 #1}\over{\partial {#2}^2}}}
\nc{\dddd}[3]{{{{\partial}^2 #1}\over
	{\partial #2 \partial #3}}}
\nc{\dder}[2]{{{d #1}\over{d #2}}}
\nc{\ddder}[2]{{{d^2 #1}\over{d {#2}^2}}}
\nc{\dddder}[3]{{d^2 #1}\over
	{d #2 d #3}}
\nc{\dx}[1]{d\,^{#1}x}
\nc{\dy}[1]{d\,^{#1}y}
\nc{\dz}[1]{d\,^{#1}z}
\nc{\dl}[1]{\frac{d\,^{#1}l}{(2\pi)^{#1}}}
\nc{\dk}[1]{\frac{d\,^{#1}k}{(2\pi)^{#1}}}
\nc{\dq}[1]{\frac{d\,^{#1}q}{(2\pi)^{#1}}}

\nc{\cc}{\mbox{$c.c.$ }}
\nc{\hc}{\mbox{$h.c.$ }}
\nc{\cf}{cf.\ }
\nc{\erfc}{{\rm erfc}}
\nc{\Tr}{{\rm Tr\,}}
\nc{\tr}{{\rm tr\,}}
\nc{\pol}{{\rm pol}}
\nc{\sign}{{\rm sign}}
\nc{\bfT}{{\bf T }}

\def\GeV{{\rm\ GeV}}

\nc{\cA}{{\cal A}}
\nc{\cB}{{\cal B}}
\nc{\cD}{{\cal D}}
\nc{\cE}{{\cal E}}
\nc{\cG}{{\cal G}}
\nc{\cH}{{\cal H}}
\nc{\cL}{{\cal L}}
\nc{\cO}{{\cal O}}
\nc{\cT}{{\cal T}}
\nc{\cN}{{\cal N}}
\nc{\rvac}[1]{|{\cal O}#1\rangle}
\nc{\lvac}[1]{\langle{\cal O}#1|}
\nc{\rvacb}[1]{|{\cal O}_\beta #1\rangle}
\nc{\lvacb}[1]{\langle{\cal O}_\beta #1 |}
\nc{\bb}{\bar{\beta}}
\nc{\bt}{\tilde{\beta}}
\nc{\ctH}{\tilde{\cal H}}
\nc{\chH}{\hat{\cal H}}
\nc{\1}{\aa}
\nc{\2}{\"{a}}
\nc{\3}{\"{o}}
\nc{\4}{\AA}
\nc{\5}{\"{A}}
\nc{\6}{\"{O}}
\nc{\al}{\alpha}
\nc{\g}{\gamma}
\nc{\Del}{\Delta}
\nc{\e}{\epsilon}
\nc{\eps}{\epsilon}
\nc{\lam}{\lambda}
\nc{\om}{\omega}
\nc{\Om}{\Omega}
\nc{\ve}{\varepsilon}
\nc{\mn}{{\mu\nu}}
\nc{\k}{\kappa}
\nc{\vp}{\varphi}

\nc{\advp}[3]{{\it  Adv.\ in\ Phys.\ }{{\bf #1} {(#2)} {#3}}}
\nc{\annp}[3]{{\it  Ann.\ Phys.\ (N.Y.)\ }{{\bf #1} {(#2)} {#3}}}
\nc{\apl}[3]{{\it  Appl. Phys. Lett. }{{\bf #1} {(#2)} {#3}}}
\nc{\apj}[3]{{\it  Ap.\ J.\ }{{\bf #1} {(#2)} {#3}}}
\nc{\apjl}[3]{{\it  Ap.\ J.\ Lett.\ }{{\bf #1} {(#2)} {#3}}}
\nc{\app}[3]{{\it Astropart.\ Phys.\ }{{\bf #1} {(#2)} {#3}}}
\nc{\cmp}[3]{{\it  Comm.\ Math.\ Phys.\ }{{ \bf #1} {(#2)} {#3}}}
\nc{\cqg}[3]{{\it  Class.\ Quant.\ Grav.\ }{{\bf #1} {(#2)} {#3}}}
\nc{\epl}[3]{{\it  Europhys.\ Lett.\ }{{\bf #1} {(#2)} {#3}}}
\nc{\ijmp}[3]{{\it Int.\ J.\ Mod.\ Phys.\ }{{\bf #1} {(#2)} {#3}}}
\nc{\ijtp}[3]{{\it Int.\ J.\ Theor.\ Phys.\ }{{\bf #1} {(#2)} {#3}}}
\nc{\jmp}[3]{{\it  J.\ Math.\ Phys.\ }{{ \bf #1} {(#2)} {#3}}}
\nc{\jpa}[3]{{\it  J.\ Phys.\ A\ }{{\bf #1} {(#2)} {#3}}}
\nc{\jpc}[3]{{\it  J.\ Phys.\ C\ }{{\bf #1} {(#2)} {#3}}}
\nc{\jap}[3]{{\it J.\ Appl.\ Phys.\ }{{\bf #1} {(#2)} {#3}}}
\nc{\jpsj}[3]{{\it J.\ Phys.\ Soc.\ Japan\ }{{\bf #1} {(#2)} {#3}}}
\nc{\lmp}[3]{{\it Lett.\ Math.\ Phys.\ }{{\bf #1} {(#2)} {#3}}}
\nc{\mpl}[3]{{\it  Mod.\ Phys.\ Lett.\ }{{\bf #1} {(#2)} {#3}}}
\nc{\ncim}[3]{{\it  Nuov.\ Cim.\ }{{\bf #1} {(#2)} {#3}}}
\nc{\np}[3]{{\it  Nucl.\ Phys.\ }{{\bf #1} {(#2)} {#3}}}
\nc{\npps}[3]{{\it  Nucl.\ Phys.\ Proc.\ Suppl.\ }{{\bf #1} {(#2)} {#3}}}
\nc{\pr}[3]{{\it Phys.\ Rev.\ }{{\bf #1} {(#2)} {#3}}}
\nc{\pra}[3]{{\it  Phys.\ Rev.\ A\ }{{\bf #1} {(#2)} {#3}}}
\nc{\prb}[3]{{\it  Phys.\ Rev.\ B\ }{{{\bf #1} {(#2)} {#3}}}}
\nc{\prc}[3]{{\it  Phys.\ Rev.\ C\ }{{\bf #1} {(#2)} {#3}}}
\nc{\prd}[3]{{\it  Phys.\ Rev.\ D\ }{{\bf #1} {(#2)} {#3}}}
\nc{\prl}[3]{{\it Phys.\ Rev.\ Lett.\ }{{\bf #1} {(#2)} {#3}}}
\nc{\pl}[3]{{\it  Phys.\ Lett.\ }{{\bf #1} {(#2)} {#3}}}
\nc{\prep}[3]{{\it Phys.\ Rep.\ }{{\bf #1} {(#2)} {#3}}}
\nc{\prsl}[3]{{\it Proc.\ R.\ Soc.\ London\ }{{\bf #1} {(#2)} {#3}}}
\nc{\ptp}[3]{{\it  Prog.\ Theor.\ Phys.\ }{{\bf #1} {(#2)} {#3}}}
\nc{\ptps}[3]{{\it  Prog\ Theor.\ Phys.\ suppl.\ }{{\bf #1} {(#2)} {#3}}}
\nc{\physa}[3]{{\it  Physica\ A\ }{{\bf #1} {(#2)} {#3}}}
\nc{\physb}[3]{{\it  Physica\ B\ }{{\bf #1} {(#2)} {#3}}}
\nc{\phys}[3]{{\it Physica\ }{{\bf #1} {(#2)} {#3}}}
\nc{\rmp}[3]{{\it  Rev.\ Mod.\ Phys.\ }{{\bf #1} {(#2)} {#3}}}
\nc{\rpp}[3]{{\it Rep.\ Prog.\ Phys.\ }{{\bf #1} {(#2)} {#3}}}
\nc{\sjnp}[3]{{\it Sov.\ J.\ Nucl.\ Phys.\ }{{\bf #1} {(#2)} {#3}}}
\nc{\spjetp}[3]{{\it Sov.\ Phys.\ JETP\ }{{\bf #1} {(#2)} {#3}}}
\nc{\yf}[3]{{\it Yad.\ Fiz.\ }{{\bf #1} {(#2)} {#3}}}
\nc{\zetp}[3]{{\it Zh.\ Eksp.\ Teor.\ Fiz.\  }{{\bf #1}  {(#2)} {#3}}}
\nc{\zp}[3]{{\it Z.\ Phys.\ }{{\bf #1} {(#2)} {#3}}}
\nc{\ibid}[3]{{\sl ibid.\ }{{\bf #1} {#2} {#3}}}
\nc{\rf}[1]{(\ref{#1})}
\nc{\nn}{\nonumber \\*}
\nc{\bfB}{\bf{B}}
\nc{\bfv}{\bf{v}}
\nc{\bfx}{\bf{x}}
\nc{\bfy}{\bf{y}}
\nc{\vx}{\vec{x}}
\nc{\vy}{\vec{y}}
\nc{\oB}{\overline{B}}
\nc{\oI}{\overline{I}}
\nc{\oR}{\overline{R}}
\nc{\rar}{\rightarrow}
\nc{\ti}{\times}
\nc{\slsh}{\hskip-5pt/}
\nc{\sm}{Standard~Model~}
\nc{\MP}{M_{\rm Pl}}
\nc{\tp}{t_{\rm Pl}}
\nc{\ave}{\bar{E}}

\nc{\eff}{{\rm eff}}
\nc{\kk}{\vek{k}}
\nc{\pp}{{\rm p}}
\nc{\ga}{g_{a\gamma}}
\nc{\vv}{\\}
\nc{\eee}{{\bf E}}
\nc{\bbb}{{\bf B}}
\nc{\qcd}{T_{\rm QCD}}
\nc{\G}{\rm \ G}
\def\vec#1{{\bf #1}}

\def\lae{\;^{<}_{\sim} \;} \def\gae{\; ^{>}_{\sim} \;} 

\begin{document}

{\title{
\vskip-2truecm{\hfill {{\small HIP-1999-47/TH\\
	\hfill \\
	}}\vskip 1truecm}
\null\vskip-75pt
{\bf  The Dynamics of Affleck-Dine Condensate Collapse}}
{\author{
{\sc  Kari Enqvist$^{1}$}\\
{\sl\small Department of Physics and Helsinki Institute of Physics,}\\ 
{\sl\small P.O. Box 9, FIN-00014 University of Helsinki, Finland}\\
{\sc and}\\
{\sc John McDonald$^{2}$}\\
{\sl\small Department of Physics and Astronomy,
University of Glasgow, Glasgow G12 8QQ, Scotland}
}
\maketitle
\begin{abstract}
\noindent

                In the MSSM, cosmological scalar field condensates formed 
along flat directions of the scalar potential (Affleck-Dine condensates) 
are typically unstable with respect to formation of  
Q-balls, a type of non-topological soliton. 
We consider the dynamical evolution of the Affleck-Dine condensate in the 
MSSM. We discuss the creation and linear growth, in F- and D-term inflation 
models, of
 the quantum seed perturbations which in the non-linear regime catalyse 
the collapse of the
 condensate to non-topological soliton lumps. 
We study numerically the evolution of the collapsing condensate lumps 
and show that the solitons initially formed are not in general Q-balls, 
but Q-axitons, 
a pseudo-breather which can have very different properties from Q-balls of the 
same charge. We calculate the energy and charge radiated from a spherically 
symmetric condensate lump as it evolves into a Q-axiton. We also discuss 
the implications for baryogenesis and dark matter. 

\end{abstract}
\vfil
\footnoterule
{\small $^1$enqvist@pcu.helsinki.fi};
{\small $^2$mcdonald@physics.gla.ac.uk}

\thispagestyle{empty}
\newpage
\setcounter{page}{1}

\section{Introduction}

       Affleck-Dine (AD) baryogenesis \cite{ad} is a natural candidate for 
the origin of the
 baryon asymmetry \cite{riotto} in the context of the MSSM and its extensions
 \cite{nilles}, with potentially important consequences for cosmology 
such as observable isocurvature perturbations \cite{bbbiso}, non-thermal 
relic neutralinos
\cite{bbbdm} and number densities of baryons and dark matter particles 
which are similar
 \cite{bbb1,bbb2}. The original picture of AD baryogenesis
was of a homogeneous squark condensate formed along an F- and D-flat direction of the
 MSSM scalar potential, in which an asymmetry is induced via non-renormalizable
soft SUSY-breaking terms. This subsequently thermalizes or decays, leaving 
the observed 
baryon asymmetry. More recently it was shown that the homogeneous AD condensate is
 typically unstable 
with respect to spatial perturbations \cite{bbb1,ks}. This is because the potential along flat
directions of the MSSM scalar potential usually deviates
from a pure $\phi^2$ potential, resulting in a negative pressure whenever the potential
is flatter than $\phi^2$. Such deviations can occur because of A-terms in the potential
 \cite{kus1}, because of a sharp change in the scalar mass term at large field values 
(gauge-mediated SUSY breaking models \cite{ks,ksd}), or because of 
radiative corrections from the gauge sector 
(conventional gravity-mediated SUSY breaking models \cite{bbb1,bbb2}). It is the latter on
which we will focus here. 
Although the main interest in unstable scalar field condensates along MSSM flat
 directions
is perhaps AD baryogenesis, we emphasize that the phenomenon can arise
 for any coherently oscillating scalar field along a flat direction of the MSSM 
scalar potential with gauge interactions. This is because the gauge corrections will cause
 the mass squared term of the flat direction to decrease with increasing scale \cite{bbb1}.
In the present view of post-inflationary SUSY cosmology, the formation of such coherently
oscillating scalar field condensates is a common and natural occurance, as a result of 
 order $H$ corrections to the SUSY breaking terms in the scalar potential \cite{h2,drt}.
 In the following we 
will generically refer to the complex scalar field as the Affleck-Dine (AD) scalar.

        Spatial "seed" perturbations in the AD condensate, arising from quantum fluctuations 
of the AD scalar during inflation, will grow and eventually go non-linear, 
forming condensate lumps \cite{bbb1,bbb2}. However, up until now there has been no clear
 picture of what 
happens to the lumps once they go non-linear. It has been assumed that they eventually 
reach their lowest energy state for a given global charge, namely a Q-ball \cite{cole}.
One aim of the present paper is to clarify the non-linear evolution of these condensate lumps. 
We will see that the condensate lumps generally do not initially 
form Q-balls, but rather a higher energy state which we refer to as a Q-axiton. 
These are somewhat similar to the original axitons \cite{axiton} (a form of pseudo-breather soliton \cite{pb}) which 
form in the case of a real axion field, but in our case they can carry a charge by virtue of
the complex nature of the AD scalar. Only in the case where the original AD condensate
is carrying a near maximal charge are the properties of the Q-axitons close to those of Q-balls of a similar charge. 

            For some of the important applications of Q-balls to 
cosmology, the ratio of the charge trapped in baryonic Q-balls (B-balls) to the total charge,
 $f_{B}$, plays an important role \cite{bbbdm,bbb2,jrev}. A second goal of this paper is to
estimate $f_{B}$ for various examples. 

        The paper is organized as follows. In section 2 we discuss the cosmology of flat
 directions in SUSY models. In section 3 we discuss the quantum seed fluctuations. In
 section 4 we discuss the linear evolution of the spatial perturbations of the AD field. 
In section 5 we study numerically the non-linear evolution of a
spherically symmetric condensate lump and estimate $f_{B}$ for various 
examples. In section 6 we discuss the implications for AD baryogenesis and 
late-decaying Q-ball cosmology. In section 7 we summarize our conclusions. 
 
\section{MSSM Flat Directions in Cosmology}

          An F- and D-flat direction of the MSSM 
with gravity-mediated SUSY breaking has a scalar potential of the form \cite{bbb1,bbb2}
\be{e1} U(\Phi) \approx (m^{2} - c H^{2})\left(1 +  K \log\left( \frac{|\Phi|^{2}}{M^{2}}
 \right) \right) |\Phi|^{2} 
+ \frac{\lambda^{2}|\Phi|^{2(d-1)}
}{M_{*}^{2(d-3)}} + \left( \frac{A_{\lambda} 
\lambda \Phi^{d}}{d M_{*}^{d-3}} + h.c.\right)    ~,\ee
where $m$ is the conventional gravity-mediated soft SUSY breaking scalar mass term
 ($m \approx 100 \GeV$), $d$ is the dimension of the non-renormalizable term in the superpotential
which lifts the flat direction, 
$c H^{2}$ gives the order $H^2$ correction to the scalar mass (with $c$ positive and
typically of the order of one for AD scalars \cite{drt}) 
and we assume that the natural scale of the 
non-renormalizable terms is $M_{*}$, where $M_{*} = M_{Pl}/8 \pi$ is the supergravity
mass scale \cite{nilles}. The A-term also receives order $H$ corrections, 
$A_{\lambda} = A_{\lambda\;o} + a_{\lambda}H$, where $A_{\lambda\;o}$ is the 
gravity-mediated soft SUSY breaking term and $a_{\lambda}$ depends on the nature of the inflation model; 
for F-term inflation $|a_{\lambda}|$ is typically of the order of one \cite{drt} whilst for
 minimal D-term inflation models it is zero \cite{kmr}. 

         The logarithmic correction to the scalar mass term, 
which occurs along flat directions with Yukawa and gauge interactions, 
is crucial for the growth of perturbations
of the AD field and the formation of Q-balls \cite{bbb1,bbb2}. This growth occurs if $K < 0$, which
 is usually the case for AD scalars with gauge interactions, 
since $K$ is dominated by gaugino corrections \cite{bbb1}. For flat directions with squark
 fields, $K \approx -(0.1-0.01)$ \cite{bbb1,bbb2}. The actual value depends on the nature of
 the flat direction; in practice we will mostly be concerned with the case of baryonic Q-balls made of squarks.
 
       If $c H^2$ is positive at the end of inflation, the AD scalar will be at a non-zero
minimum of its potential. Once 
$H \lae m$, the AD scalar will begin to coherently oscillate around its minimum
at zero and the A-term will induce a global charge asymmetry in the condensate
 \cite{ad,drt,jrev}. 
The resulting asymmetry in the Universe at present is then proportional to the reheating
 temperature after inflation ($T_{R}$), 
\be{ea1} \eta_{B} = \frac{2 \pi n_{B} T_{R}}{H^2 M_{Pl}^2}      ~,\ee
where $\eta_{B}$ is the present baryon to entropy ratio and $n_{B}$ and $H$ are the
 baryon charge density and the expansion rate when the asymmetry is formed ($H \approx 
m$) \cite{bbb2}. We see that, because of the dependence on $T_{R}$, the baryon asymmetry in AD baryogenesis 
cannot be determined independently of the inflation model; in practice, we can use the 
present baryon asymmetry to {\it fix} the reheating temperature and so determine the 
background cosmology. In general, the Universe will be matter dominated
 by coherently oscillating inflatons when the B asymmetry forms at $H \approx m$, since
 the thermal gravitino upper bound on $T_{R}$, 
$T_{R} \lae 10^{9} \GeV$ \cite{grav}, implies that the value of $H$ when the inflaton
 matter domination period ends at $T_{R}$ is less than 1 GeV.   

\section{Seed Perturbations} 
       
\subsection{Growth of Perturbations}

         Just like for the inflaton, there will be a spectrum of spatial perturbations induced in the AD scalar field
by quantum fluctuations during inflation \cite{bbb1}. For fields of mass $m \lae H$, 
the scalar fields have fluctuations on leaving the horizon of magnitude
\be{sp1}   \delta \phi \approx \frac{H}{2 \pi}              ~.\ee
In general, the equations of motion for perturbations about the minimum of the potential
 have the form, 
\be{sp2} \delta \ddot{\phi} + 3 H \delta \dot{\phi} - \nabla^2 \delta \phi = - k H^2 \delta \phi       ~,\ee
where $k$ is determined by the parameters $c$, $|a_{\lambda}|$ and $d$ of the AD potential. 
For perturbations larger than the horizon, the spatial derivative can be ignored, since the
 perturbation can be treated as homogeneous on sub-horizon scales. 
During inflation (with $H = H_{I} =$ constant), the solution of Eq.(4) for the amplitude of
 larger than horizon perturbations is
\be{sp4} \delta \phi \propto a^{-Re(\sigma)}    \;\;\;\;\;\; ; \;\;   \sigma = \frac{1}{2} 
\left[ 3 - \sqrt{9 - 4 k } \right]         ~.\ee
$k > 9/4$ gives a damped and oscillating solution for $\delta \phi$, for which the 
suppression of the amplitude of the perturbation by the expansion of the Universe is
 maximal, whilst $k < 9/4$ gives a purely
damped solution. In general,  $Re(\sigma) \leq 3/2$.
 During the matter dominated period following inflation (with $H \propto 
a^{-3/2}$), the solution for the amplitude of $\delta \phi$ is 
\be{sp3} \delta \phi \propto a^{-Re(\eta)}    \;\;\;\;\;\; ; \;\;   \eta = \frac{1}{2} 
\left[ \frac{3}{2} - \sqrt{\frac{9}{4} - 4 k } \right]         ~.\ee
$k > 9/16$ gives a damped and oscillating solution for $\delta \phi$, with maximal amplitude suppression, 
whilst $k < 9/16$ gives a purely damped solution. In general, $Re(\eta) \leq 3/4$.  Thus
 when the perturbation re-enters the horizon during matter domination, the amplitude is given by
\be{sp5} \delta \phi_{hor}  = 
\left( \frac{a_{e}}{a_{hor}} \right)^{Re(\eta)} 
 \frac{e^{-Re(\sigma)  \Delta N_{e}(\lambda)} H_{I}}{2 \pi}      ~,\ee 
where $\Delta N_{e}(\lambda)$ is the number of e-foldings before the end of inflation 
at which a perturbation of scale $\lambda$ exits the horizon, $a_{e}$ is the scale factor 
at the end of inflation and $a_{hor}$ is the scale factor when the perturbation re-enters the
horizon. Once inside the horizon, the spatial derivatives become important. For $ 
\frac{\vec{k}^2}{a^2} \gae H^2$ and $H \gae m$, the spatial derivative dominates the potential 
term and $\delta \phi \propto a^{-1}$. Thus, writing in terms of the scale 
$\lambda$ at $H$ during matter domination, the perturbation spectrum once the 
peturbation is within the horizon is given by
\be{sp6} \delta \phi (\lambda) \approx {(H^2H_{I}\lambda^3)^x \over 2 \pi \lambda}\; ; 
~~x = 1 -\frac{2}{3} Re \left( \eta \right) - \frac{1}{3} Re\left(\sigma\right)   ~.\ee
For the case with the largest suppression of the amplitude ($k > 9/4$), $Re(\sigma) = 
2 Re(\eta) = 3/2$, so that $x=0$ and the 
resulting spectrum is very simple\begin{footnote}{This differs from the 
spectrum given in \cite{bbb1}, which assumed a constant mass term throughout for the 
perturbation.}\end{footnote}
\be{sp7}  \delta \phi (\lambda) \approx \frac{1}{2 \pi \lambda}            ~.\ee
An important feature of the spectrum in what follows is that the perturbation is usually
 larger for smaller values of $\lambda$. 

\subsection{Perturbations of the Affleck-Dine Field}

        The actual value of $k$ is determined by the form of the AD scalar potential. Let
 $\Phi = (\phi_{1} + i \phi_{2})/\sqrt{2}$ and let $k_{1}$ and $k_{2}$ be the values 
of $k$ for $\delta \phi_{1}$ and $\delta \phi_{2}$. In general, choosing the phase of $\Phi$
 such that $\lambda A_{\lambda}$ is real and negative, the minimum of the potential 
is at $\phi_{2} = 0$ and $\phi_{1} = \phi_{1\;m}$, with
\be{se1} \phi_{1\;m} = \left(\frac{c}{\alpha}\right)^{\frac{1}{2\left(d-2\right)}}
H^{\frac{1}{d-2}}  \;\;\; ; \; \alpha = \frac{\lambda^{2} (d-1)}{2^{d-2} M_{*}^{2(d-3)}}
      ~,\ee
where for simplicity we are considering the A-term along the 
$\phi_{1}$ direction to be negligible when minimizing; in general it results in a correction of
 order one to the 
$\phi_{1}$ minimum. During inflation, $H$ is constant and, on perturbing about the
 $(\phi_{1\;m}, 0)$ minimum, we obtain equations of the form of Eq.(4) with 
\be{se5a}    k_{1} = c (2 d-4) ~\ee
for $\delta \phi_{1}$ and 
\be{se5b}    k_{2} = |a_{\lambda}| \left(d-1\right)^{1/2} c^{1/2}      ~\ee
for $\delta \phi_{2}$.

            During the matter dominated period the minimum of the AD potential 
is time-dependent. As a result, when we perturb about $\phi_{1 \;m}$ by $\Delta \phi_{1}$,
 we obtain 
\be{se2} \Delta \ddot{\phi}_{1} + 3 H \Delta \dot{\phi}_{1} 
= \frac{9}{4} \frac{\left(d-3\right
)}{\left(d-2\right)^{2}} H^{2} \phi_{1\;m} 
- c (2d-4) H^{2} \Delta \phi_{1}     ~.\ee
Thus $\phi_{1}$ will not oscillate about $\phi_{1\;m}$, but 
rather about a larger value given by 
\be{se3} \phi_{1}(t) = \left(1 + \frac{9 \left(d-3\right)}{8 c \left(d-2\right)^{3}} \right) \phi_{1\;m}        ~.\ee
Perturbing ($\phi_{1}$, $\phi_{2}$) about the ($\phi_{1}(t)$, 0) effective minimum then
 gives equations of the form of Eq.(4) with
\be{se4a}    k_{1} = c (2 d-4) + \frac{9}{4}
 \frac{\left(d-3\right) \left(2d-3\right)}{\left(d-2\right)^{2} }
 ~\ee
and
\be{se4b}    k_{2} = \frac{9}{4}
\frac{\left(d-3\right)}{\left(d-2\right)^{2}}
+ |a_{\lambda}|\left(d-1\right)^{1/2} c^{1/2} \left(1 + \frac{9 \left(d-3\right)}{8 c \left(d-2\right)^{2}} \right)      ~.\ee
If $a_{\lambda} = 0$, as in D-term inflation models, then the solution for the evolution of
 $\delta \phi_{2}$ is simply 
$\delta \phi_{2} \propto \phi_{1}(t)$. This is easily understood, as in this limit there is 
no potential along the $\theta$ direction of the AD field and so $\delta \theta 
= \delta \phi_{2}/\phi_{1}$ is constant. 

        From this we see that, in general, the evolution of the $\phi_{1}$ and $\phi_{2}$
 perturbations at $H \gae m$ can be different, depending on $c$, $d$ and $a_{\lambda}$.
 However, for $c$ and $|a_{\lambda}| \approx 1$, as expected in F-term inflation models, 
it is likely that $k_{i} > 9/4$ and so the amplitude of  $\delta \phi_{1}$ and $\delta \phi_{2}$
 will be equally (i.e. maximally) suppressed throughout.
In this case $\delta \phi_{1} \approx \delta \phi_{2}$ at $H \approx m$. We will 
focus on this case in the following. For $|a_{\lambda}| = 0$, as expected in D-term inflation
 models, $\delta \phi_{2} \propto \phi_{1}$ and so $\delta \theta$ is constant. In this case
 $\delta \phi_{2}$ can be much larger than $\delta \phi_{1}$ at $H \approx m$. 

          To discuss the evolution of the perturbations at $H \lae m$, we need to know the
 initial spectrum of spatial perturbations of the AD field when the condensate and baryon
 asymmetry forms at $H \approx m$. The directions $\phi_{1}$ and $\phi_{2}$ 
refer to the real and imaginary directions as determined by the phase of the A-term. 
However, the phase of the A-term changes as $H$ becomes smaller than $m$ and 
$A_{\lambda\;o}$ comes to dominate $a_{\lambda}H$. Therefore,
\newline (i) For the $|a_{\lambda}| 
\approx 1$ case, corresponding to F-term inflation, the effect is to rotate $\phi_{1}$ and
 $\phi_{2}$ such that, if we consider
 the relative phase of $A_{\lambda\;o}$ and $a_{\lambda}H$ to be of the order of one
(corresponding to the case of a condensate with amplitude $\phi_{1} \approx \phi_{2}$ at
 $H \lae m$), then regardless of whether $\delta \phi_{1}$ or $\delta \phi_{2}$ dominates at
 $H \gae m$, 
$\delta \phi_{1}$ will be approximately equal to $\delta \phi_{2}$ once $H \lae m$. 
\newline (ii) If $|a_{\lambda}| \approx 1$ but  
the relative phase is much smaller than 1 (corresponding to the case of a
 condensate with amplitude $\phi_{1} \gg \phi_{2}$ at $H \lae m$), then it is possible for
 $\delta \phi_{2}/\phi_{2}$ to be much larger than 
$\delta \phi_{1}/\phi_{1}$ at $H \lae m$, for example if $\delta \phi_{1} \approx \delta
 \phi_{2}$ at $H \gae m$ and $\phi_{1} \gg \phi_{2}$ at $H \lae m$. 
\newline (iii) For the case with $a_{\lambda} = 0$, corresponding to D-term inflation,
the initial phase of the AD field will initially be of order one relative to that of
 $A_{\lambda \;o}$, so that, as with (i), $\delta \phi_{1} \approx \delta \phi_{2}$ and
 $\phi_{1} \approx \phi_{2}$ once $H \lae m$.

      Thus F-term inflation models can produce condensates which have different
values for $\delta \phi_{i}/\phi_{i}$ ($i=1,2$) at $H \approx m$, 
whereas minimal D-term inflation models always produce roughly equal values for 
$\delta \phi_{i}/\phi_{i}$. 

\section{Linear Evolution}

         Because of the logarithmic correction to the AD scalar potential, there 
will be an attractive force between the condensate scalars which causes the spatial perturbations to grow.
The inflationary perturbations are known to have a small amplitude so that 
the initial growth will take place in the linear regime. 
Condensate fragmentation and collapse to soliton lumps happens in the non-linear regime,
 for which the linear evolution provides calculable initial conditions. Therefore we will first
 consider in detail the linear evolution of the perturbations.
The homogeneous AD condensate can be characterized by the charge asymmetry 
it carries. The condensate is generally a sum of two real oscillating scalar fields. 
The maximum possible charge corresponding to a given maximum amplitude of the 
AD scalar occurs when the combined oscillation corresponds to a circle in the ($\phi_{1}, \phi_{2}$) 
plane. We will refer to this as a maximally charged (MAX) condensate. This is roughly
 expected to occur in the case of F-term inflation models with an 
order one CP violating phase and in the case of minimal D-term inflation models. The condensate
 can also carry a less than maximal charge, as in the case of F-term inflation models with a
 small CP violating phase, in which case it will describe an ellipse in the ($\phi_{1}, \phi_{2}$) plane. 

\subsection{MAX Condensate} 

            A solution for the linear evolution of spatial perturbations of the condensate 
has been given previously for the case of a MAX condensate \cite{bbb2,ks}, which we
 refer to as the Kusenko-Shaposhnikov (KS) solution \cite{ks}.
It assumes a solution of the form $\phi = \phi(t) + \delta \phi(x,t)$ and 
$\theta = \theta(t) + \delta \theta(x, t)$, where the homogeneous MAX condensate 
is described by
\be{le1} \Phi = \frac{\phi(t)}{\sqrt{2}}e^{i \theta(t)}       ~,\ee
 with $\phi(t) = (a_{o}/a)^{3/2}\phi_{o}$ ($a$ is the scale factor) and
$\dot{\theta}(t)^{2} \approx m^{2}$, up to corrections of order $Km^{2}$. 
(As it is true for most models, we will assume that $|K|$ is small compared with 1.)
 The KS solution also requires that 
$\delta \phi(x,t)$ and $\delta \theta (x,t)$  initially satisfies 
\be{ks1}    \delta \theta_{i} \approx \left(\frac{\delta \phi}{\phi}\right)_{i}    ~.\ee
As discussed in the previous section, for a MAX condesate we expect $\delta
 \phi_{1}/\phi_{1} \approx \delta \phi_{2}/\phi_{2}$, so that this condition is 
satisfied. 
The solution of the linear perturbation equations then has the form \cite{bbb2}
\be{le2} \delta \phi \approx \left(\frac{a_{o}}{a}\right)^{3/2}
 \delta \phi_{o}\;  exp \left( \int dt \left(\frac{1}{2} \frac{\vec{k}^{2}}{a^{2}}
 \frac{|K| m^{2}}{\dot{\theta}(t)^{2}} \right)^{1/2} \right)  
e^{i\vec{k}.
\vec{x}}   
  ~\ee
and 
 \be{le3} \delta \theta \approx \delta \theta_{i}\;  exp \left( \int dt \left(\frac{1}{2} 
\frac{\vec{k}^{2}}{a^{2}}
 \frac{|K| m^{2}}{\dot{\theta}(t)^{2}} \right)^{1/2} \right)   
e^{i\vec{k}\cdot
\vec{x}}       ~.\ee
These apply if $ \left| \vec{k}^{2}/a^{2} \right| \lae | 2K m^{2}|   $,              
$H^{2}$ is small compared with $m^{2}$ and $|K| \ll 1$. 
If the first condition is not satisfied, the gradient energy of the perturbations
 produces a positive pressure larger than the negative pressure due to the attractive
 force from the logarithmic term, 
preventing the growth of the perturbations. 

        For the case of a matter dominated Universe, the exponential growth factor is 
\be{le4} \int dt \left(\frac{1}{2} \frac{\vec{k}^{2}}{a^{2}}
 \frac{|K| m^{2}}{\dot{\theta}(t)^{2}} \right)^{1/2} = \frac{2}{H} \left(
\frac{|K|}{2} \frac{\vec{k}^{2}}{a^{2}}\right)^{1/2}      ~,\ee 
where we take the scale factor when the AD oscillations begin to be equal to 1.
The largest growth factor will correspond to the largest value of $\vec{k}^{2}$ for which
 growth
can occur, $\vec{k}^{2}/a^{2} \approx 2 |K| m^{2}$. Thus the value of $H$ at which the
first perturbation goes non-linear is \cite{bbb2}
\be{le5} H_{i} \approx \frac{2 |K| m}{\alpha(\lambda)}      ~,\ee
with
\be{le6} \alpha(\lambda) = - log \left( \frac{\delta \phi_{o}(\lambda)}{\phi_{o}}\right)    ~,\ee
where $\phi_{o}$ is the value of $\phi$ when the condensate oscillations begin at 
$H \approx m$. A typical value of $\alpha(\lambda)$ for the spectrum Eq.(9) (e.g. for
 $d=6$)
 is $\alpha(\lambda) \approx 
30$. The initial non-linear region has a radius $\lambda_{i}$ at $H_{i}$ given by
\be{le7} \lambda_{i} \approx \frac{\pi}{|2 K|^{1/2} m}    ~.\ee
Thus for a MAX condensate, when the condensate is just going non-linear, the amplitude
 of the scalar field is given by 
\be{le7a} \phi(\vec{x},t) = \phi(t) + \delta \phi(\vec{x},t) \approx \phi(t)(1+cos(\vec{k}.\vec{x}))      ~,\ee
where $|\vec{k}|$ is given by the wavelength which first goes non-linear, $\lambda_{i}$.
 This will provide the initial condition for the numerical study of the
 non-linear regime.

\subsection{Non-MAX Condensate}

            For the case of a non-MAX condensate, we cannot directly use the KS solution. 
However, it is likely that the initial radius and the time at which the spatial perturbations
 initially go 
non-linear will roughly be the same as for the MAX condensate. 
We will assume that $\delta \phi_{1} \approx \delta \phi_{2}$, as is true if both scalar
perturbations are
 maximally suppressed when $H \gae m$. The equations of motion for $\phi_{i}$ (i=1,2) are 
 given by
\be{ks2} \ddot{\phi}_{i} + 3H \phi_{i}- \nabla^{2} \phi_{i} = -m^{2}(1+K)\phi_{i} -K m^2 \phi_{i}
log\left(\frac{\phi_{1}^{2} + \phi_{2}^{2}}{\phi_{o}^{2}} \right)   ~.\ee
For $\phi_{1} \gg \phi_{2}$, the equation for $\phi_{1}$ will be similar to the case of the 
MAX condensate, except the $\phi_{1}^{2} + \phi_{2}^{2}$ will be oscillating in time 
rather than constant. Thus the equation for the growth of perturbations in 
$\phi_{1}$ will be similar to the MAX condensate and the $\phi_{1}$ perturbations 
will go non-linear ($\delta \phi_{1}/\phi_{1} \gae 1$) roughly as in the case of the 
MAX condensate. The equation for perturbation in $\phi_{2}$ will, however, be different, 
since the {\it log} term is dominated by $\phi_{1}^{2}$. The condition for the $\phi_{2}$ 
equation to go non-linear is then that $\delta \phi_{2} \gae \phi_{1}$.  
We expect the rate of growth of $\delta \phi_{2}$ to be no larger than for $\delta \phi_{1}$,
 since the perturbations grow due to the attractive interaction between the scalars due to
 the {\it log} term, and the interaction of $\phi_{2}$ with $log(\phi_{1}^{2})$
is the same as that of $\phi_{1}$. So if $\delta \phi_{1} \approx \delta \phi_{2}$ initially then
 we expect $\delta \phi_{2} \lae \delta \phi_{1}$ throughout. Therefore non-linearity will 
occur only once $\delta \phi_{1}$ goes non-linear, at which point the condensate will begin
 to fragment to condensate lumps. In general, the charge density of the initial 
non-linear lumps will be essentially the same as that of the original homogeneous
 condensate. 

\section{Non-linear Evolution and $f_{B}$}

\subsection{Initial Lump}

        In order to study the non-linear evolution, we will consider the evolution of a 
single condensate lump. We can neglect the expansion of the Universe, since 
at the time when the spatial perturbations go non-linear, the expansion rate, $H_{i}$, is
 small compared with the dynamical mass scale in the equations of motion governing the
 growth of the lumps, $|K|^{1/2}m$, so that the $3 H \dot{\phi}$ terms in the equations of 
motion are negligible compared with the other terms. 
From the discussion of Sect. 4, when the spatial perturbation is
just going non-linear ($\delta \phi \approx \phi$), the AD field for the MAX condensate can
 be written as 
\be{g5} \Phi(x,t) \approx  \frac{Ae^{imt}}{\sqrt{2}}(1 + cos(\vec{k}.\vec{x}) )    ~,\ee
where $A/\sqrt{2}$ is the amplitude of the coherent oscillations when the field 
first goes non-linear.
In terms of $\Phi = (\phi_{1} + i \phi_{2})/\sqrt{2}$, the initial non-linear perturbation 
is described by
\be{g6}  \phi_{1}(\vec{x},t) = A cos(mt) (1 + cos(\vec{k}.\vec{x}) )       ~\ee
\be{g7}  \phi_{2}(\vec{x},t) = A sin(mt) (1 + cos(\vec{k}.\vec{x}) )       ~.\ee
This can be thought of as a 'lattice' of adjacent condensate lumps. In order to 
gain some insight into the evolution of these lumps, we will consider a single spherically
 symmetric lump. Such lumps are described in general by
\be{g8}  \phi_{1}(r,t) = A cos(mt) (1 + cos(\pi r/2 r_{0}) )       ~\ee
\be{g9}  \phi_{2}(r,t) = B sin(mt) (1 + cos(\pi r/2 r_{0}) )       ~,\ee
for $r \leq 2r_{0}$ and by $\phi_{1,2} = 0$ otherwise. The initial radius of the lump 
is $2 r_{0}$, where $r_{0} =  \pi/(\sqrt{2} |K|^{1/2} m)$. The MAX condensate lump 
corresponds to $A = B$, whilst the non-MAX lump has $A > B$. 
The corresponding energy and charge densities (with unit charge for the scalars) are 
\be{g12} 
\rho = |\dot{\Phi}|^2 + |\nabla\Phi|^2 + V(|\Phi|) ~
\ee
and
\be{e13} q  = \phi_{1} \dot{\phi}_{2} - \dot{\phi}_{1} \phi_{2}        
~,\ee
where
\be{g12a} V(|\Phi|) = m^2 |\Phi|^{2} \left(1 + K  log \left( 
\frac{|\Phi|^{2}}{|\Phi|_{o}^{2}} \right) \right)  ~.
\ee
The total energy and charge in a fixed volume are given respectively
by
\be{QE}
E=4\pi\int_V drr^2\rho~~,~Q=4\pi\int_V drr^2 q~\sim AB,
\ee
with, for the initial MAX condensate lump of Eqs.(30, 31), 
\be{Qmax}
Q=Q_{\rm max}\equiv 132.2A^2|K|^{-3/2}m^{-2} ~.\ee
The non-maximally charged lump with $B < A$ 
then corresponds to $Q/Q_{\rm max} = B/A < 1$.

\subsection{Numerical Solution}

        The scalar field equations of motion are given by Eq.(26) with $H=0$.
We have solved these equations numerically for the case of
the spherically symmetric lump. 
We choose $A=1$ and set $m=100$ GeV when needed.
To avoid the singularity at $|\phi|=0$,
where the one-loop logarithmic correction alone no longer is valid,
we have introduced a cut-off $\delta$ by letting $(\phi_{1}^{2} + \phi_{2}^{2})
/\phi_{o}^{2}
\to (\phi_{1}^{2} + \phi_{2}^{2})/\phi_{o}^{2}+\delta$; we have verified that
the value of the small cut-off does not significantly affect the solutions.
With the initial lump radius $2r_0$ we have considered a spatial sphere of 
radius $8r_0$, at the boundary of which the outflowing waves are damped
by hand. This is to prevent reflected waves bouncing back onto the lump,
a situation which would be realistic only if the average lump distance
were extremely small.

The behaviour of the solutions depends on
$K$, and to a greater extent on $Q/Q_{\rm max}$. In general, the 
condensate lump pulsates while charge is flowing out until
the lump reaches a (quasi-)equilibrium pseudo-breather configuration\begin{footnote}{We
 distinguish between spatial pulsations of the lump as it collapses and breathing, which is
 essentially the coherent oscillation of the AD field within the lump.}\end{footnote} (in
 which the 
lump pulsates with only a small difference between the maximum and minimum 
field amplitudes), as seen in Fig. 1a. We refer to this state as a Q-axiton. For
 non-maximal condensates, it is very different from the 
lowest energy configuration for a given charge, namely a Q-ball. The Q-ball 
is essentially made entirely of charge, with $E \approx mQ$ (neglecting the small 
binding energy per charge) \cite{bbb1,bbb2}. For the Q-axiton, in which the attractive force
 between the scalars is balanced by the gradient pressure of the scalar field, the energy per
 unit charge can be much larger than $m$; indeed, the Q-axiton exists even if $Q=0$. 
Only for a more or less maximally charged Q-axiton are the properties close to that of the 
corresponding Q-ball. We believe that the Q-axiton should eventually evolve to the lower 
energy Q-ball state\begin{footnote}{There are no absolutely stable spherically symmetric
 breather solutions in $3+1$ flat space \cite{axiton}.}\end{footnote}, although this is not
 apparent from our numerical results and we cannot rule out the possibility that
 the Q-axiton may be very long-lived. 
\begin{figure}
\leavevmode
\centering
\vspace*{90mm} 
\includegraphics{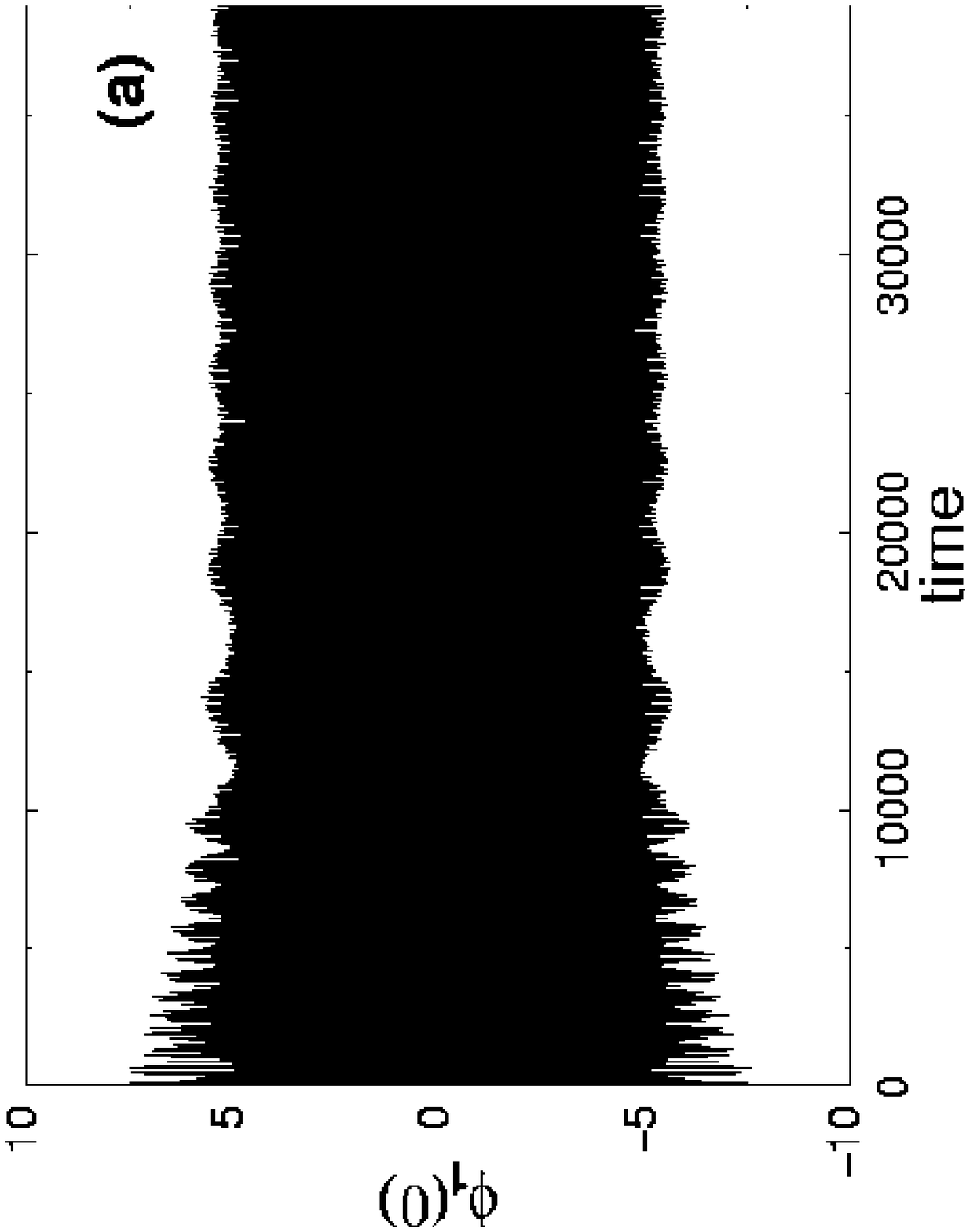}
\includegraphics{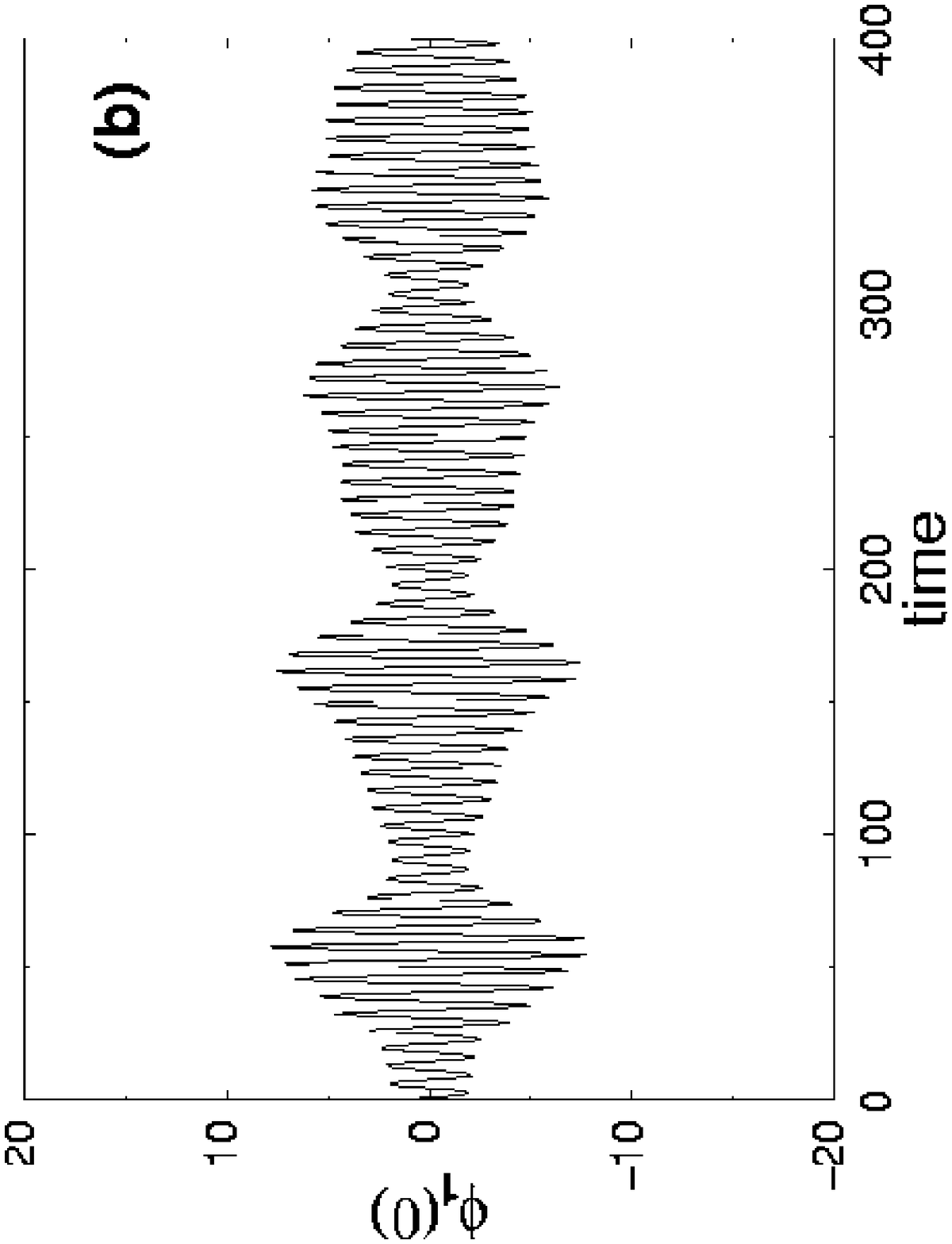}
\includegraphics{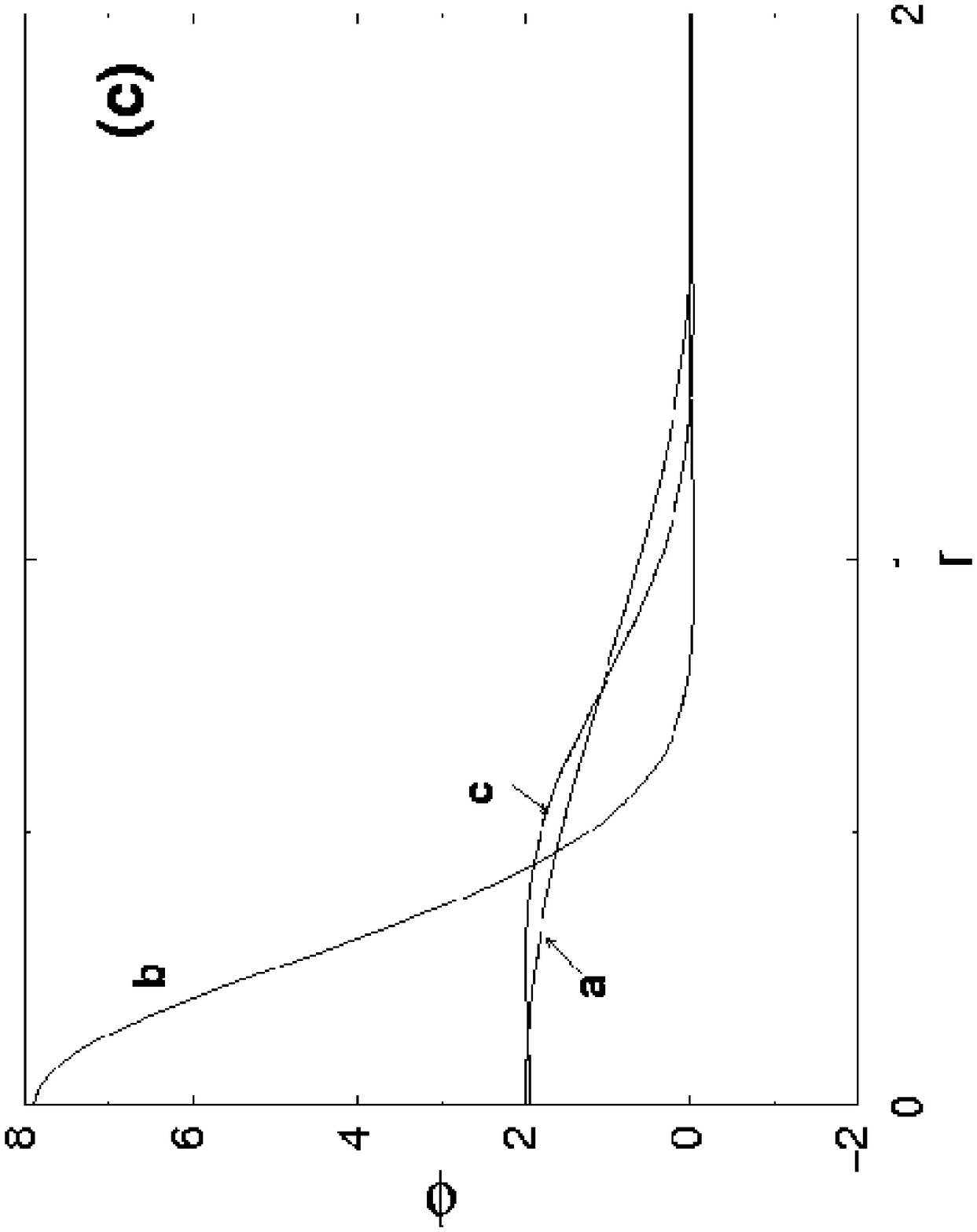}
\includegraphics{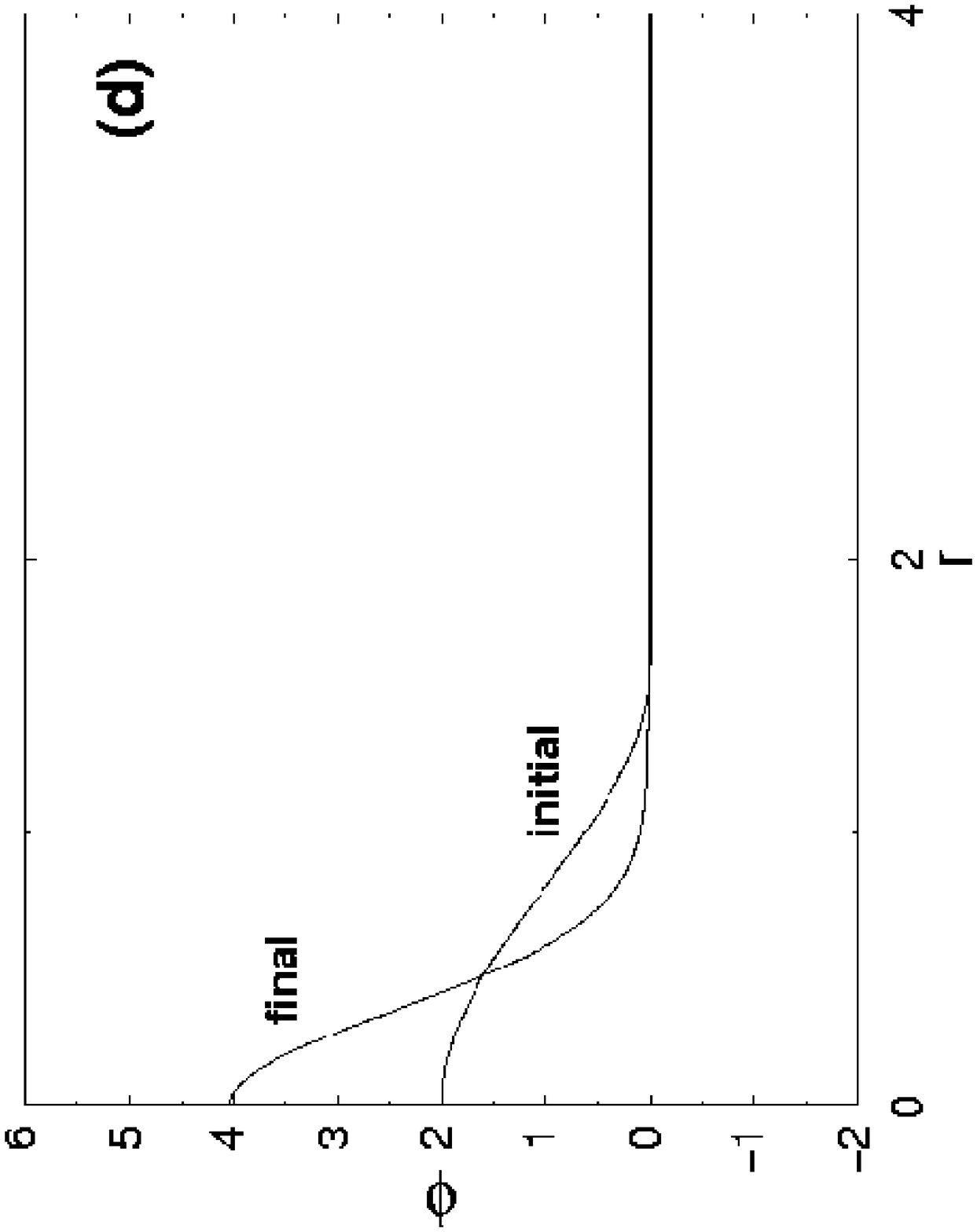}
\caption{(a) The time evolution of the field amplitude at the origin,
$\phi_1(0)$; (b) a detail of the early time evolution, showing
the pulsation cycles modulating the coherent oscillations (time is in
units of $1/m$); (c) the first pulsation of the condensate lump:
a) initial lump; b) maximal lump; c) minimal lump; (d) the final
equilibrium Q-axiton compared with the initial lump ($r$ is in units
of $r_0$). In all cases $K=-0.05$ and $Q=Q_{\rm max}$.} 
\label{kuva1}       
\end{figure} 
Compared with the
natural time scale $t=1/m$, the time taken to reach the Q-axiton state 
is long, of
the order of $1600/m$ for the
case $K = -0.05$ and $Q=Q_{\rm max}$, which is about five expansion
 times\begin{footnote}{Expansion can, however, be neglected in numerical solution for the
 Q-axiton itself, as noted previously.}\end{footnote} ($H_{i}^{-1} \approx 300/m$). This is
 shown in Fig. 1, where we display the oscillation of the
 field amplitude at the origin as a function of time, as well as the
spatial profile of the whole lump as it pulsates. Fig. 1a shows the rapid pulsations
gradually settling towards the equilibrium by emitting scalar field waves. Fig 1b 
shows the first four pulsations of the lump; the high frequency oscillations 
modulated by the pulsations are the coherent oscillations of the AD field. 
The lump keeps its shape
during the whole time while pulsating and slowly evolving to the final 
configuration, as depicted in Figs. 1c and 1d, which show the maximum amplitudes 
at different stages of the pulsations.
In the case with $Q = Q_{max}$,
$\phi_1$ and $\phi_2$ have almost identical time evolution, the only difference
being that there is a phase difference of $\pi/2$ between 
the $\phi_1$ and $\phi_2$ oscillations, as can easily be understood from the symmetry of
 the initial conditions and scalar field equations of motion.

       In Fig.\ 2 we show the time evolution of the charge and the energy
of the whole configuration, integrated out to the distance
$8r_0$.  At first charge and energy flows out of the volume,
but the slow approach to equilibrium can also readily be seen,
with the axiton lump starting with initial charge $Q=1.18$ and
stabilizing to $Q\simeq 1.09$. $Q$ and $E$ are nearly proportional, as they should be for
 the case with $Q_{max}$ 
and $\phi_{1}$ and $\phi_{2}$ different only by a phase $\pi/2$. (In fact, $E/Q$ decreases
 from $106 \GeV$ initially to $98 \GeV$, due to the increased binding energy of
 the Q-axiton state.)
We see that for the case of the MAX condensate the ratio of $E$ to $Q$ is close to
 $m = 100 \GeV$ throughout, showing that the Q-axiton formed in this case is essentially a Q-ball. 
\begin{figure}
\leavevmode
\centering
\vspace*{60mm} 
\includegraphics{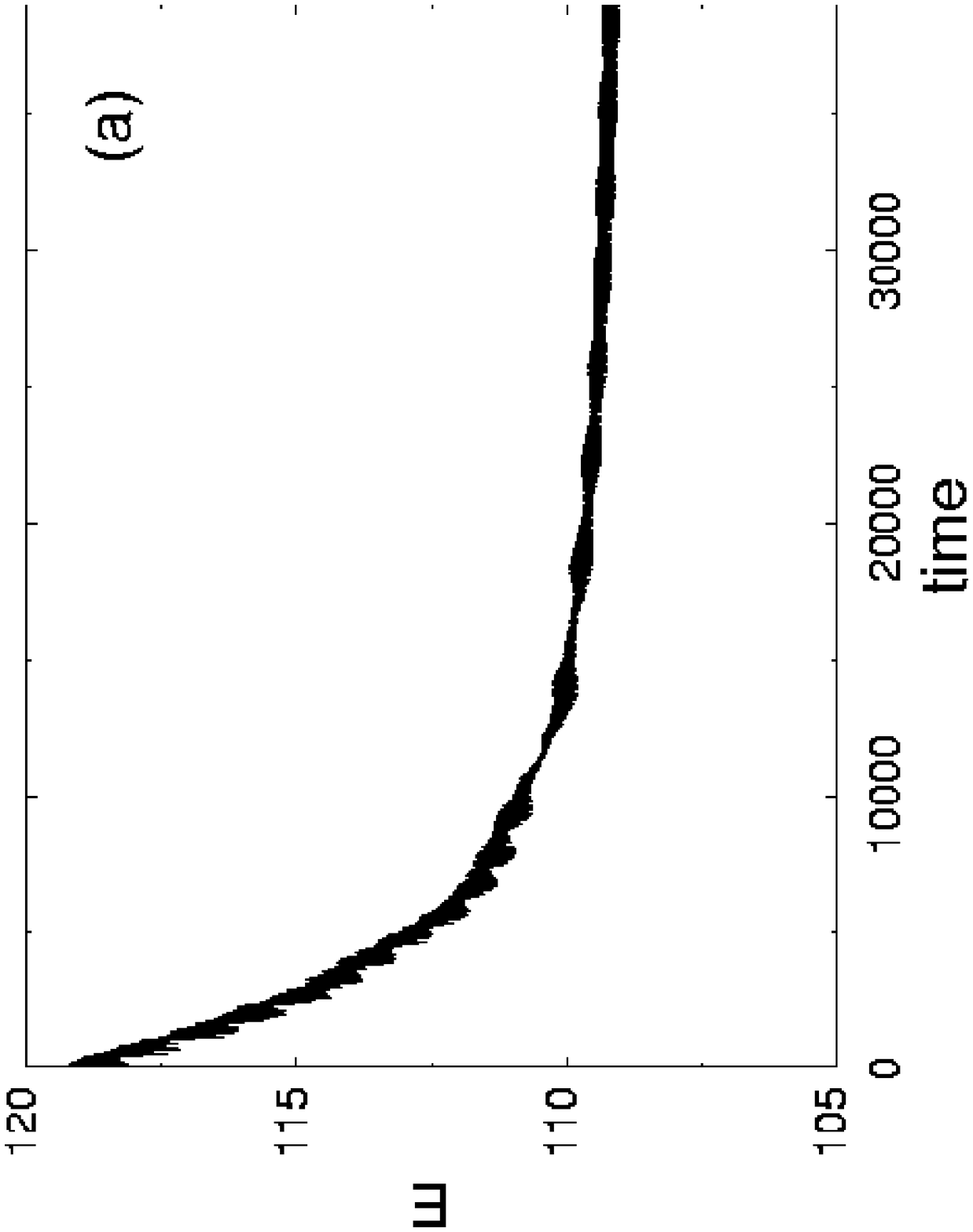}
\includegraphics{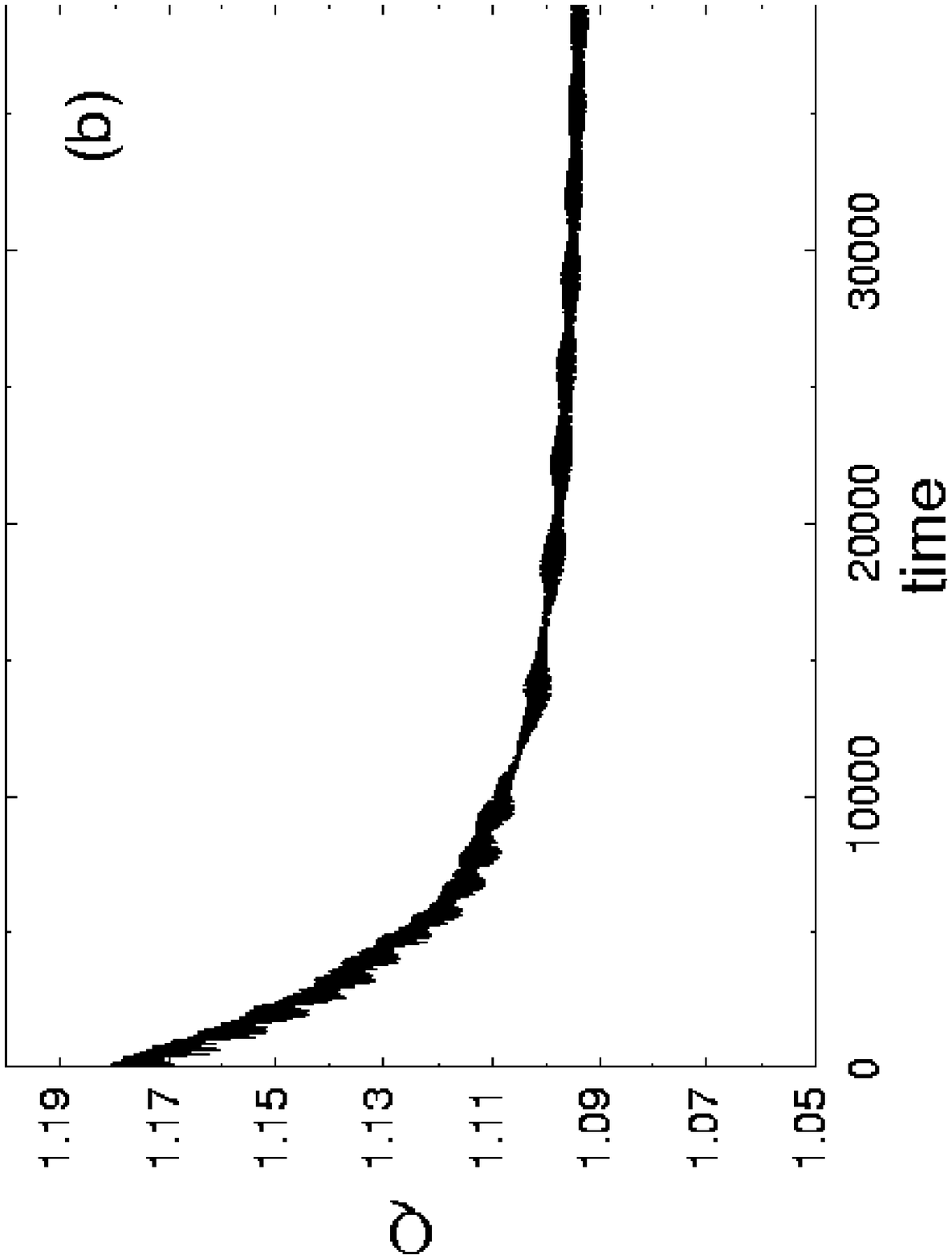}
\caption{(a) The time evolution (in
units of $1/m$) of the Q-axiton energy $E$ and (b)
charge $Q$ for  $K=-0.05$ and initial $Q=Q_{\rm max}$.} 
\label{kuva2}       
\end{figure} 

In Fig.\ 3 we show the field amplitude and lump oscillations for
the non-maximally charged case $Q=0.1Q_{\rm max}$. Now the 
squared $\phi_1$
and $\phi_2$ amplitudes do not follow a circle as in the
maximally charged case, but rather a precessing
ellipse (Fig. 3b). The qualitative behaviour is similar to the maximally
charged case in that while $E$ and $Q$ at first escape the fiducial
volume, at later times the configuration reaches a quasi-equilibrium
state (Fig. 4). The initial fluctuation in $Q$ is a numerical artifact; in general, our 
numerical solution for the evolution of the charge of the lump becomes 
less accurate for smaller $Q/Q_{max}$. The ratio of $E$ to $Q$ of the Q-axiton in this
 case is approximately $600 \GeV$, much larger than for the corresponding Q-ball.
\begin{figure}
\leavevmode
\centering
\vspace*{95mm} 
\includegraphics{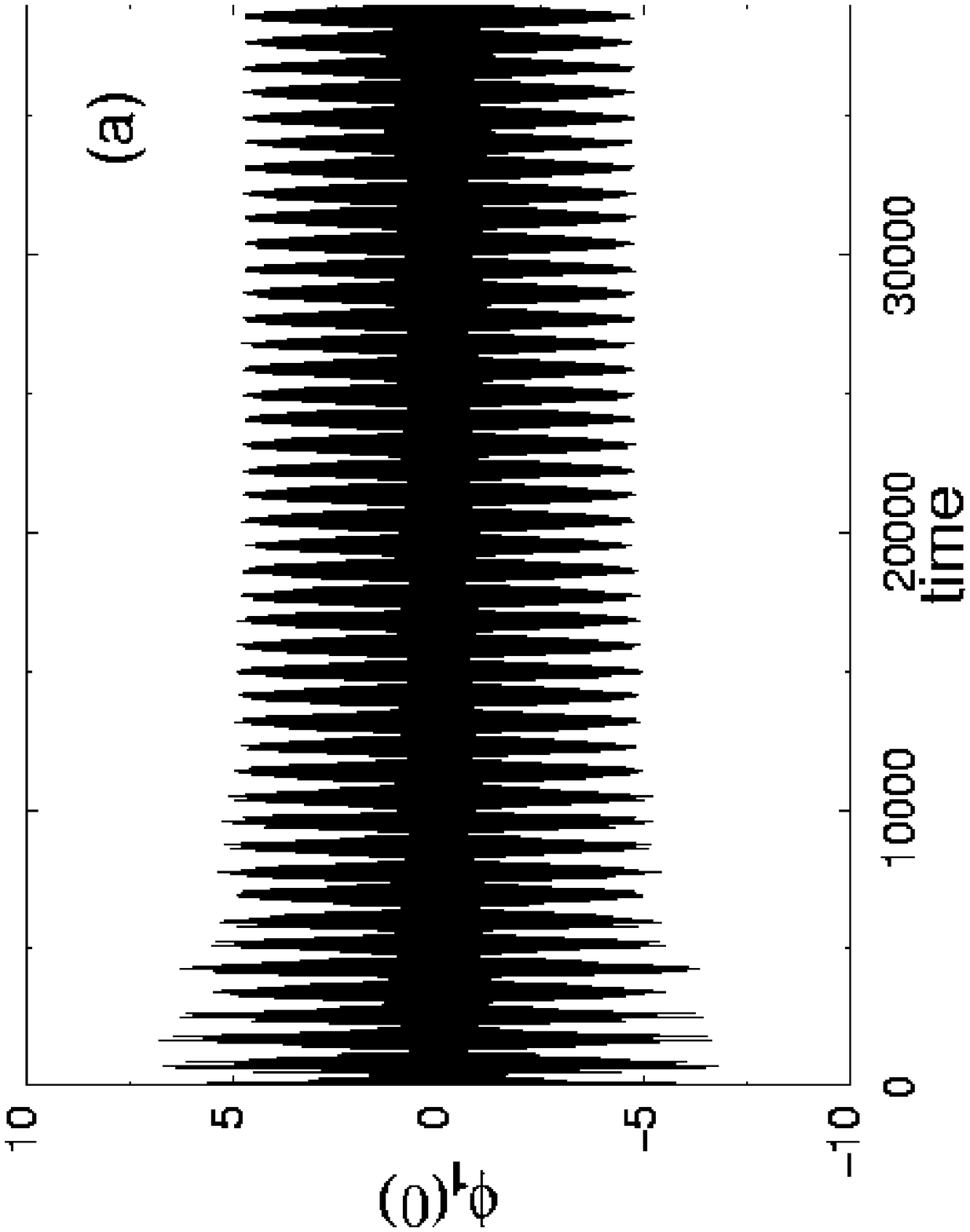}
\includegraphics{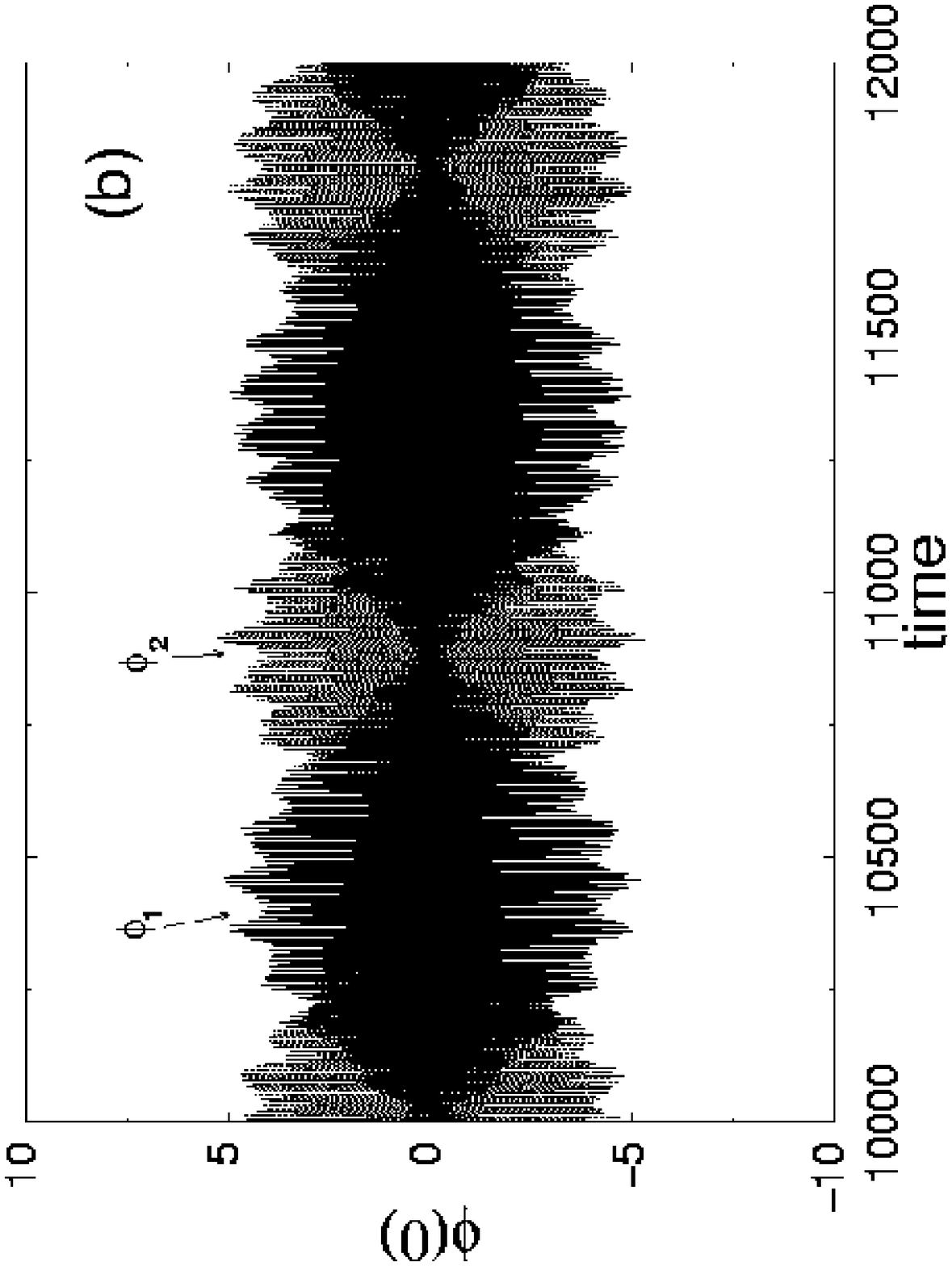}
\includegraphics{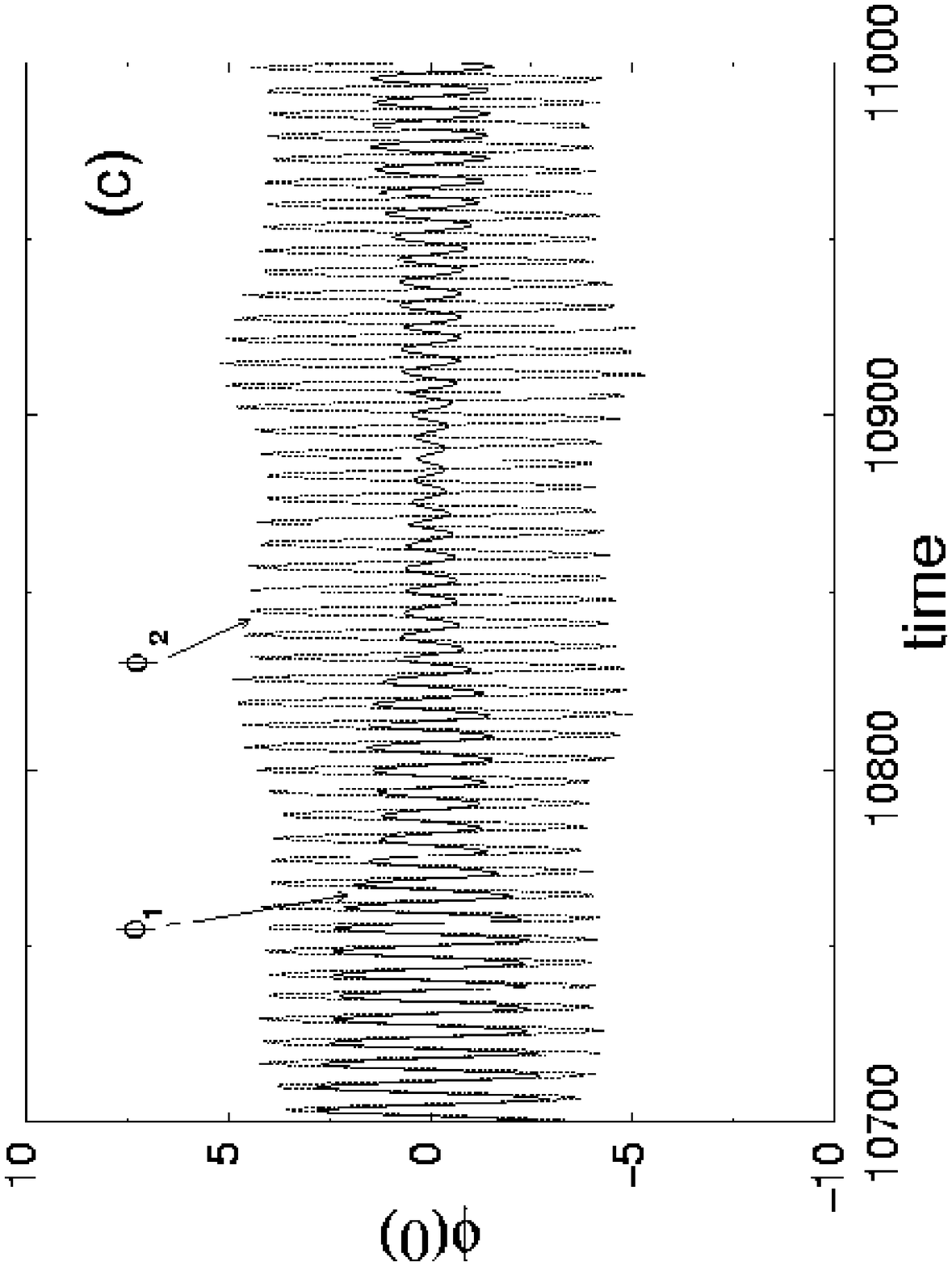}
\caption{(a) The time evolution of the field amplitude at the origin,
$\phi_1(0)$; (b) a detail of the time evolution, showing
both $\phi_1(0)$ and $\phi_2(0)$; (c) a further detail.
In all the cases $K=-0.05$ and $Q=0.1Q_{\rm max}$
while time is in
units of $1/m$} 
\label{kuva3}       
\end{figure} 

\begin{figure}
\leavevmode
\centering
\vspace*{60mm} 
\includegraphics{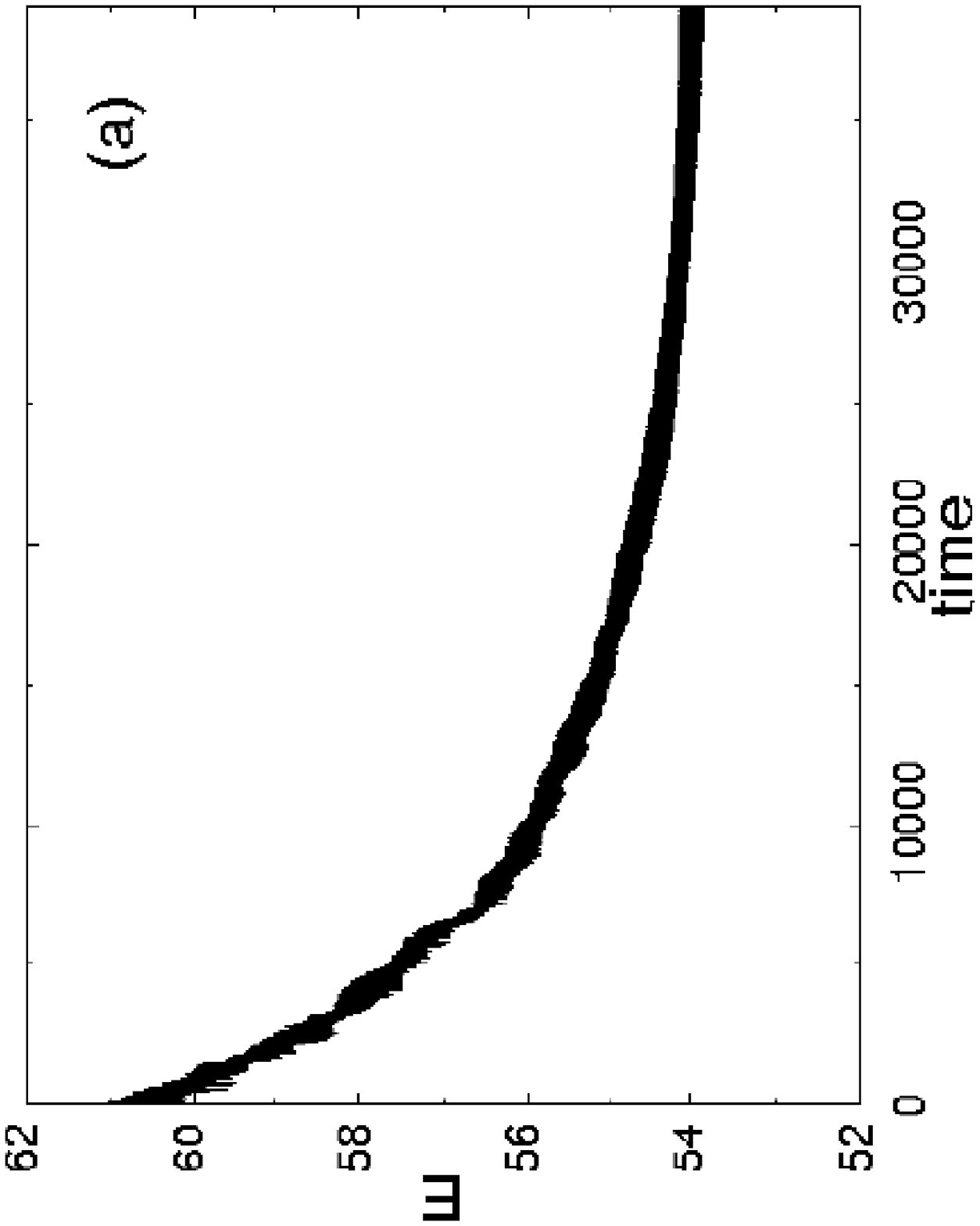}
\includegraphics{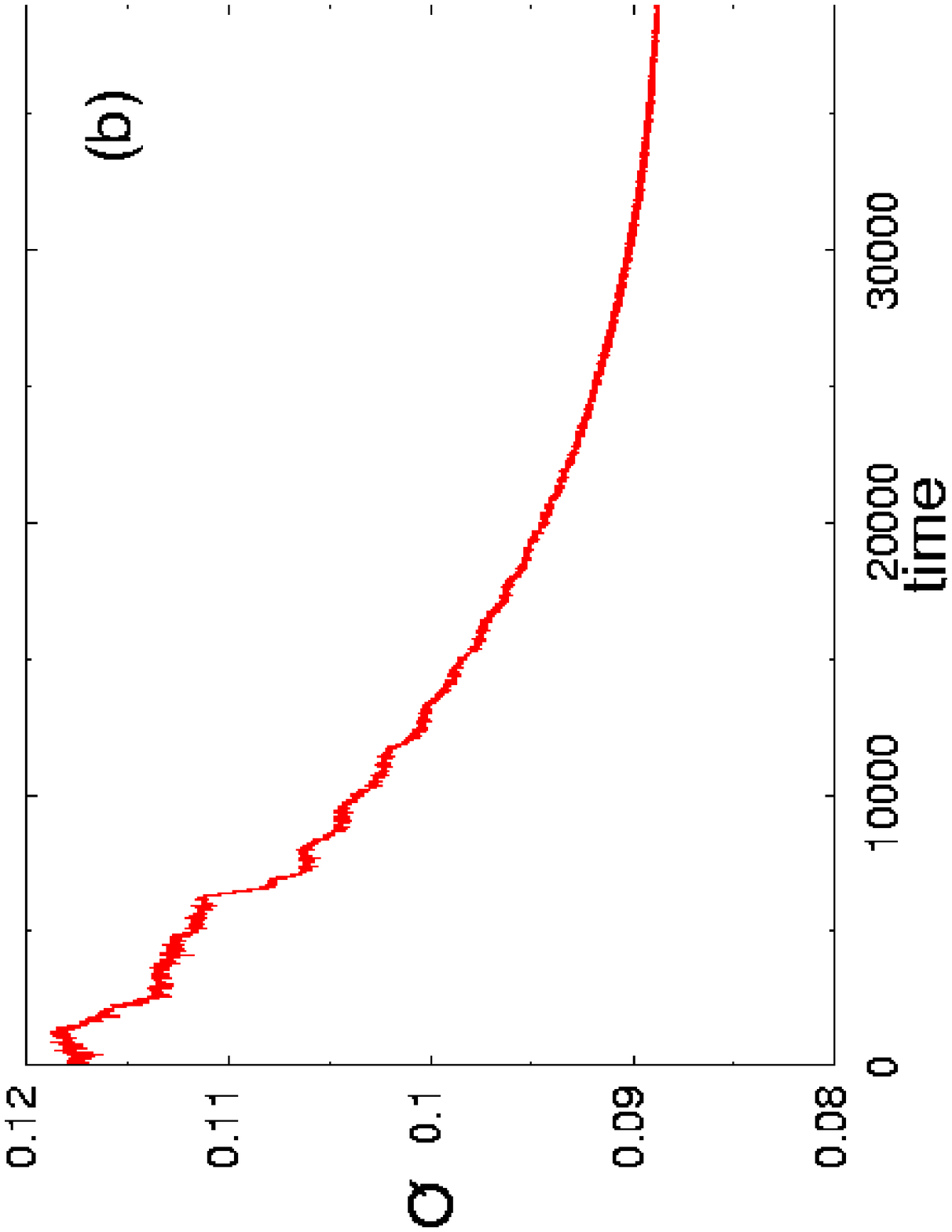}
\caption{(a) The time evolution (in
units of $1/m$) of the Q-axiton energy $E$ and (b)
charge $Q$ for  $K=-0.05$ and initial 
$Q=0.1Q_{\rm max}$.} 
\label{kuva4}       
\end{figure} 

\subsection{Values of $f_{B}$}

            We have also calculated $f_{B}$, the ratio of the charge in the equilibrium 
Q-axiton state to the charge of the initial condensate lump. The results as a function of 
 $Q$ and $K$ are given in Table 1, from which we see that $f_{B}$
 decreases for smaller values of $Q$ and for larger vales of $|K|$. 
\begin{center} {\bf Table 1. Values of $\bf f_{B}$} \end{center} 
\begin{center}
\begin{tabular}{|c|c|c|}          \hline
$K$ & $Q/Q_{max}$ & $f_B$  \\ \hline
$-0.1$ & $1$ &  $0.89$
 \\
$-0.05$ & $1$ &  $0.92$
 \\
$-0.01$ & $1$ &  $0.94$
 \\
\hline
$-0.1$ & $0.1$ &  $0.64$
 \\
$-0.05$ & $0.1$ &  $0.74$
 \\
$-0.01$ & $0.1$ &  $0.83$
 \\
\hline
$-0.1$ & $0.01$ &  $\sim 0.3$
 \\
$-0.05$ & $0.01$ &  $\sim 0.5$
 \\
$-0.01$ & $0.01$ &  $\sim 0.6$
 \\
\hline
\end{tabular}
\end{center} 

               In all this we have considered a single spherically symmetric lump isolated 
in space, compared with the real case in which there is a  
lattice of closely spaced lumps which are in general not spherically symmetric. It is 
beyond the scope of the present paper to study this more realistic case, but we note that 
the rate at which the lumps radiate charge and energy is very slow. By the time the lumps
 have radiated their charge and reached the Q-axiton state, they will have been
 substantially pulled apart by expansion, so that the lumps are eventually isolated.
 Therefore the results for $f_{B}$ for the spherically symmetric lump are likely to be a
 reasonable estimate of the 
values in the realistic case. Indeed, since we might expect non-spherically symmetric 
lumps to be less efficient in reaching their quasi-equilibrium configurations, the values of
 $f_{B}$ for the spherically symmetric lump may well be an underestimate of the realistic
 value. We will assume in the following that the Q-axitons eventually evolve to lower
 energy Q-balls of the same charge.
 To do this they must lose their excess energy, presumably by a very slow radiation of
 scalar field waves. However, the full evolution of the Q-axiton to the final Q-ball requires
 a much more ambitious numerical analysis than that attempted here.

\section{Consequences for Affleck-Dine Baryogenesis}

          We have previously discussed the possibility that Q-balls can decay at a very
low temperature, $T_{d} \lae 1 \GeV$.  Such low decay temperatures 
naturally occur for the case of $d=6$ 
AD baryogenesis, which has a reheating temperature typically around 1 GeV
 \cite{bbb2,jrev}. Late decaying Q-balls ("B-ball baryogenesis"\cite{bbb1,bbb2}) can both
 protect the baryon asymmetry from the effects of L 
violating interactions (such as arise with Majorana neutrino masses) and allow for 
an understanding of why the number density of baryons and dark matter particles are 
within an order of magnitude of each other when the dark matter particles have masses 
of the order of $m_{W}$. In the case where the similarity of the number
densities is explained by the late decay of Q-balls to lightest SUSY particles (LSPs)
 and baryons,
the value of $f_{B}$ immediately gives us the mass of the LSP.
The ratio of the number densities is given by \cite{bbbdm,bbb2}  
\be{h1}    \frac{n_{B}}{n_{DM}} = \frac{1}{3 f_{B}}          ~\ee
The observed range of ratios is \cite{jrev}
\be{h2}     \left(\frac{n_{B}}{n_{DM}}\right)_{obs} = (1.5-7.3) 
\frac{m_{LSP}}{m_{W}}          ~.\ee
To be consistent with present experimental limits on MSSM neutralinos, 
$m_{\chi} \gae 30 \GeV$ \cite{nulim}, we would require that $f_{B} \lae 0.6$. 
From Table 1 we see that this rules out MSSM neutralinos
as the LSP for $K$ in the range $-0.1$ to $-0.01$ and $Q/Q_{max} \geq 0.1$. However, 
for $Q/Q_{max} = 0.01$ it is probable that, for sufficiently large $|K|$, light MSSM neutralinos 
from Q-ball decay can be consistent with experimental bounds. 
 Our numerical results become less accurate for smaller values of $Q$ and larger $|K|$,
 but we expect that MSSM neutralinos up to $60 \GeV$ can be consistent 
with experimental limits for $K = -0.1$
 and $Q/Q_{max} = 0.01$. It is interesting to note that a recent dark matter search
 experiment has suggested
the existence of a candidate of mass $59 \GeV$ \cite{wimp}. It should be kept in mind that
 in the realistic non-spherically
 symmetric case $f_{B}$ could well be smaller, so allowing for 
larger neutralino masses. On the
 other hand, if $f_{B} \gae 0.6$, then either we must allow  
the neutralinos to annihilate after Q-ball decay, so breaking the direct connection between
 the number densities of baryons and dark matter particles \cite{bbbdm,bbb2}, or we must
 go beyond the MSSM. A light singlino in the
 next-to-minimal SUSY Standard Model (NMSSM) would be an interesting possibility for
 the LSP in this case \cite{nmssm}. 

       This all assumes that the Q-axitons evolve into the corresponding Q-balls. Should the
 Q-axitons be very long-lived, their cosmology, for the case of a non-MAX condensate, 
could be substantially different from that of the Q-balls .

\section{Conclusions}

             We have considered the origin and linear evolution, in the context of SUSY F- and
 D-term inflation models, of spatial perturbations of 
an Affleck-Dine condensate and the collapse of a spherically symmetric condensate lump 
of the type expected to arise from the fragmentation of the condensate. 
We find that for a typical range of parameters, 30-90 $\%$ of the charge of
 the collapsing lump ends up in the final quasi-equilibrium state ($f_{B} \approx 0.3-0.9$),
 with the 
remainder being radiated in the form of ripples 
of scalar field. The quasi-equilibrium state is, in general, not a Q-ball, but a higher energy
 pseudo-breather state, a Q-axiton. This
physically approximates a Q-ball only for the case of an initial condensate which is nearly
maximally charged. 
Assuming that the Q-axiton evolves into the corresponding Q-ball, for the case where dark
 matter and baryons originate directly from late-decaying Q-balls, so explaining the 
similarity of the number densities of baryons and dark matter particles, $f_{B} \lae 0.6$
 allows for a range of neutralino masses consistent with experimental 
constraints on MSSM neutralinos.
Such values can occur if the condensate is not maximally charged.
 $f_{B} \gae 0.6$, on the
 other hand, requires $m_{LSP} \lae 30 \GeV$, which rules out MSSM neutralinos 
coming directly from late-decaying Q-balls as dark matter, although an NMSSM singlino
 could be consistent with experimental constraints. 
The numerical results presented here are based on the evolution of a single, isolated, 
spherically symmetric condensate lump. To study Affleck-Dine condensate collapse for the
 realistic case, with many condensate lumps of different shapes and sizes, and to 
study the evolution of the Q-axitons into Q-balls, we would require a much more
ambitious numerical analysis than that presented here. We hope to develop this in the
 future. 

\subsection*{Acknowledgements}   This work has been supported by the
 Academy of Finland and by the UK PPARC.

\end{document}